\documentclass[letterpaper,twocolumn,10pt]{article}
\usepackage{usenix,epsfig,endnotes}
\usepackage{booktabs}
\usepackage{xspace}
\usepackage{multirow}
\usepackage{listings}
\usepackage{amsmath}
\usepackage{subcaption}
\usepackage{wrapfig}
\usepackage{pifont}
\usepackage{xurl}

\newcommand{\cmark}{\ding{51}}%
\newcommand{\xmark}{\ding{55}}%
            
\usepackage{xcolor}

\definecolor{codegreen}{rgb}{0,0.6,0}
\definecolor{codegray}{rgb}{0.5,0.5,0.5}
\definecolor{codepurple}{rgb}{0.58,0,0.82}
\definecolor{backcolour}{rgb}{0.95,0.95,0.92}

\lstdefinestyle{mystyle}{
    backgroundcolor=\color{backcolour},   
    commentstyle=\color{codegreen},
    keywordstyle=\color{magenta},
    numberstyle=\tiny\color{codegray},
    stringstyle=\color{codepurple},
    basicstyle=\ttfamily\footnotesize,
    breakatwhitespace=false,         
    breaklines=true,                 
    captionpos=b,                    
    keepspaces=true,                 
    numbers=left,                    
    numbersep=5pt,                  
    showspaces=false,                
    showstringspaces=false,
    showtabs=false,                  
    tabsize=2
}
\usepackage[linesnumbered, ruled]{algorithm2e}
\usepackage[misc]{ifsym}
\SetKwFunction{FunctionName}{FunctionName}
\SetKwProg{Fn}{Function}{:}{}
\SetKwRepeat{Do}{do}{while}%

\newcommand{\sysname}{numaPTE\xspace}

\newcommand{\revised}[1]{{\textcolor{black}{#1}}}
\usepackage[affil-it]{authblk}
\makeatletter
\renewcommand\AB@affilsepx{\quad\quad \protect\Affilfont}
\def\thanks#1{\protected@xdef\@thanks{\@thanks
        \protect\footnotetext{#1}}}
\makeatother

\begin{document}

\date{}

\title{\Large \bf numaPTE: Managing Page-Tables and TLBs on NUMA Systems}

\author[1]{Bin Gao}
\author[1]{Qingxuan Kang}
\author[1]{Hao-Wei Tee}
\author[2]{Kyle Timothy Ng Chu}
\author[3]{Alireza Sanaee}
\author[1]{Djordje Jevdjic}

\affil[1]{\normalsize{National University of Singapore}}
\affil[2]{\normalsize{Horizon Quantum Computing}}
\affil[3]{\normalsize{Queen Mary University of London}}

\maketitle

\subsection*{Abstract}
Memory management operations that modify page-tables, typically performed during memory allocation/deallocation, are infamous for their poor performance in highly threaded applications, largely due to process-wide TLB shootdowns that the OS must issue due to the lack of hardware support for TLB coherence. We study these operations in NUMA settings, where we observe up to 40x overhead for basic operations such as \texttt{munmap} or \texttt{mprotect}. The overhead further increases if page-table replication is used, where complete coherent copies of the page-tables are maintained across all NUMA nodes. While eager system-wide replication is extremely effective at localizing page-table reads during address translation, we find that it creates additional penalties upon any page-table changes due to the need to maintain all replicas coherent.

In this paper, we propose a novel page-table management mechanism, called \emph{\sysname}, to enable transparent, on-demand, and partial page-table replication across NUMA nodes in order to perform address translation locally, while avoiding the overheads and scalability issues of system-wide full page-table replication. We then show that \sysname's precise knowledge of page-table sharers can be leveraged to significantly reduce the number of TLB shootdowns issued upon any memory-management operation. As a result, \sysname not only avoids replication-related slowdowns, but also provides significant speedup over the baseline on memory allocation/deallocation \revised{and access control operations}. We implement \sysname in Linux on \texttt{x86\_64}, evaluate it on 4- and 8-socket systems, and show that \sysname achieves the full benefits of eager page-table replication on a wide range of applications, while also achieving \revised{a 12\% and 36\% runtime improvement on Webserver and Memcached respectively due to a significant reduction in TLB shootdowns.}

\section{Introduction}
\label{sec:intro}

Translation Lookaside Buffers (TLBs) are tiny  hardware structures that cache recently used virtual-to-physical address mappings from the page-tables. Since TLBs must be accessed before every memory operation, their latency is critical to the performance of the entire system, which is why they must be small and tightly integrated into every CPU core. Given that every core has its own TLB that independently caches page-table entries (PTE), a mechanism that enforces coherence of PTEs across all TLBs that cache them is required for correctness. Unfortunately, most modern multi-core architectures do not provide hardware support for TLB coherence, but instead provide privileged instructions for invalidation of TLB entries, which the operating system (OS) calls upon any change to the page-tables. Such instructions can invalidate TLB entries only on the core that executes them; to invalidate TLBs on all other cores, OSes use an expensive, IPI-based mechanism to send an interrupt to each core individually due to the lack of support for flexible multi-cast delivery~\cite{oskin:tlb, kumar:latr}, in a procedure that's called \emph{TLB Shootdown}. Furthermore, when a PTE changes, the OS lacks the information about which TLB in the system currently caches the modified PTE, and thus must send shootdowns indiscriminately to all cores that currently run a thread of the same process, and it must do so synchronously for correctness, causing delays of several microseconds or even tens of microseconds on big machines~\cite{kumar:latr}.

\begin{figure*}[htpb]
    \begin{minipage}[htpb]{0.5\linewidth}
        \centering
        \includegraphics[width=3.2in]{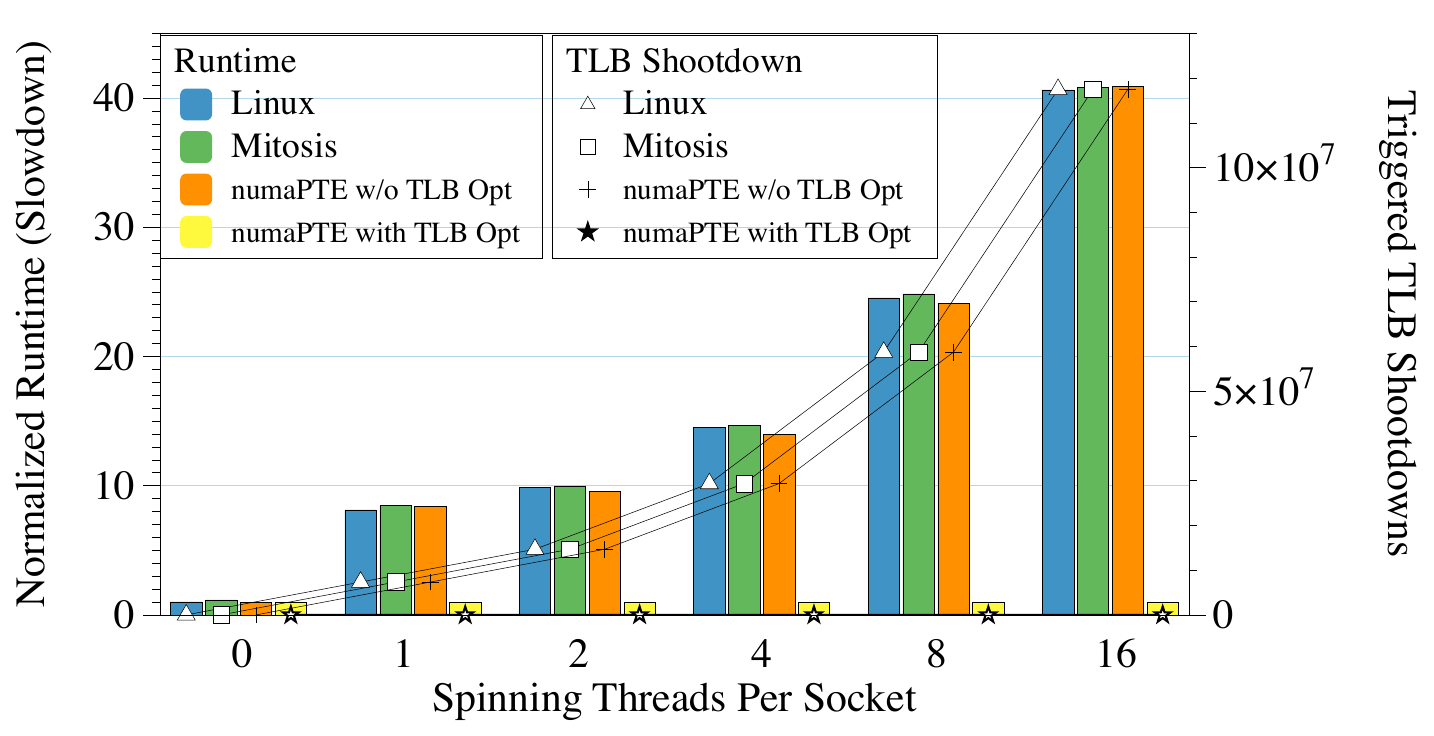} 
        \begin{center}
            \revised{a) \sysname improvement over baseline Linux(v4.17) and Mitosis}
        \end{center}
    \end{minipage}
\hfill
    \begin{minipage}[htpb]{0.5\linewidth}
        \centering
        \includegraphics[width=3.2in]{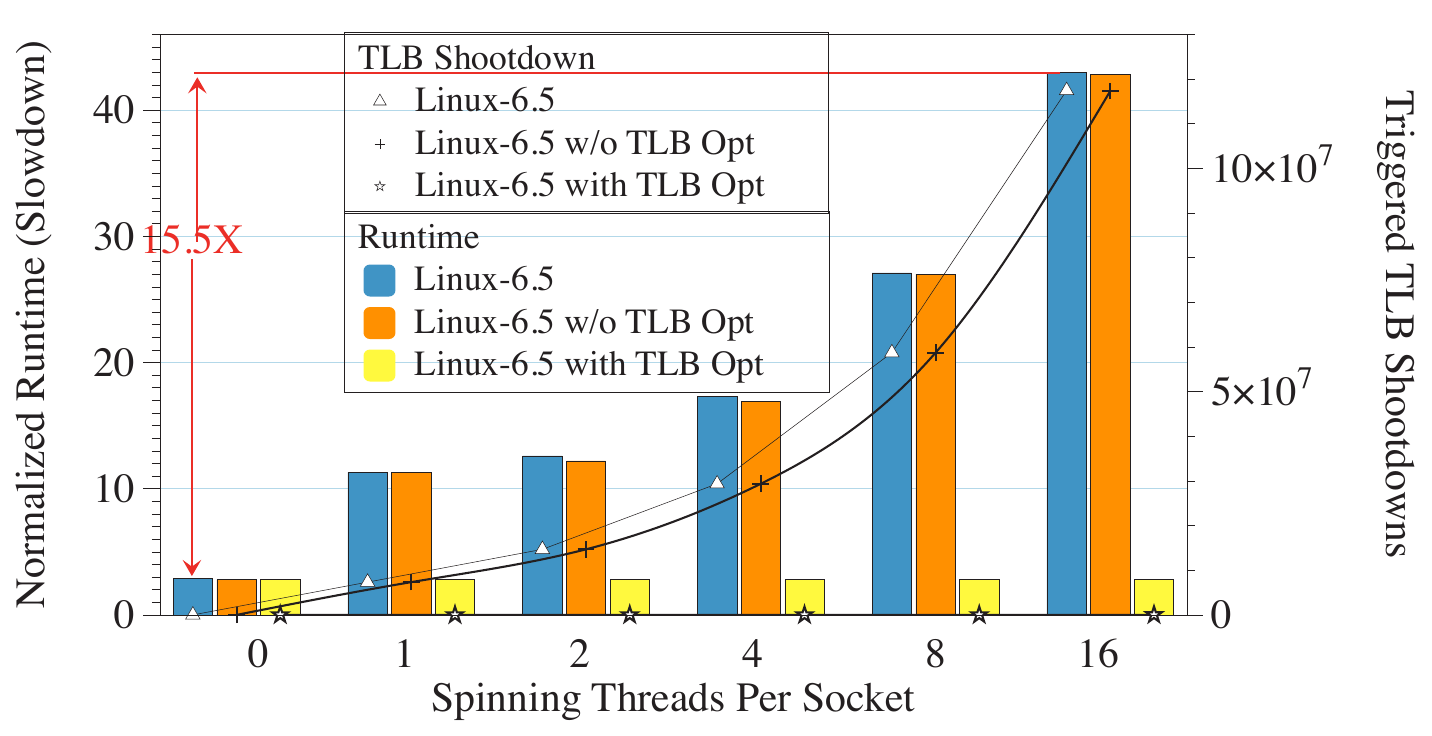}
        \begin{center}
            \revised{b) \sysname improvement over Linux(v6.5.7)}
        \end{center}
    \end{minipage}
    \caption{\revised{Impact of page-table replication, and the TLB shootdown optimization on \texttt{mprotect}. \sysname reduces the run time slow down by up to 40x, and mitigates the TLB shootdown overhead by leveraging the information about page tables on each socket. All values in both plots are normalized to the baseline Linux v4.17 without replication.}}
    \label{fig:mprot1}

\end{figure*}

\begin{figure}[h]
        \includegraphics[width=1\columnwidth]{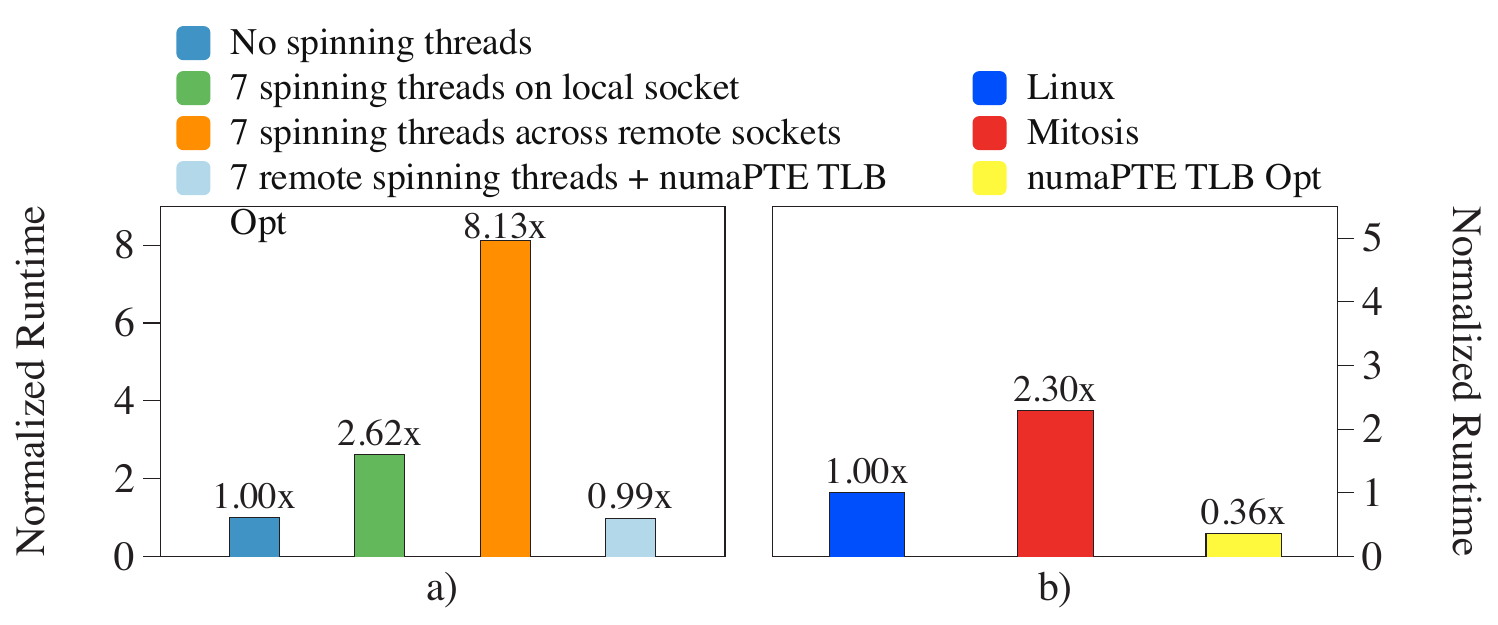}
        \caption{a) The slowdown Linux on \texttt{mprotect} with local threads spinning on the local vs. remote sockets, b) The slowdown of Mitosis and \sysname over Linux when the range of \texttt{mprotect} is 512KB; note that Mitosis sees a slowdown, while \sysname experiences a speedup.}
        \label{fig:mprot2}
\end{figure}

The TLB coherence operations particularly affect multi-socket and multi-node systems, which data centers have shifted towards in order to continue to scale up the CPU performance and memory capacity in the post-Moore's law era. These systems connect multiple CPUs to multiple memory modules in a NUMA fashion. Unfortunately, this has a dramatic impact on TLB coherence. To illustrate, Figure~\ref{fig:mprot1} shows the performance of \texttt{mprotect}, a Linux syscall that changes the permission bits in page-table, when called for a 4KB page on an 8-socket machine. This experiment is performed using the testbed explained in Section~\ref{subsec:hwspec}. In this experiment, a single thread runs \texttt{mprotect} in a loop, repeatedly flipping a single bit in a single PTE. Additionally, we run a varying number of spinning threads on every socket. \revised{The spinning threads mimic the ideal behavior of scale-out workloads where numerous threads perform independent computations with limited data-sharing and synchronizations among them.}
As Figure~\ref{fig:mprot1} shows, the TLB shootdowns cause a 40x performance degradation for \texttt{mprotect} on the baseline Linux v4.17 when spinning threads are added to other sockets. This problem persists in newer kernels, such as v6.5.7, with up to 15.5x performance degradation as shown in Figure~\ref{fig:mprot1}b. Note that the new kernel nominally shows better scalability, as it degrades performance by only 15.5x, vs. 40x with v4.17. However, this is due to the fact that the baseline \texttt{mprotect} performance (without spinning threads) is about 3x worse compared to Linux v4.17.\footnote{Linux v4.17 performs better in absolute terms for all our workloads, and unless otherwise mentioned, all results we show are based on v4.17.} Also note that the impact of shootdowns sent to spinning threads on remote sockets is significantly higher compared to the shootdowns sent to threads that are spinning on the same socket as where \texttt{mprotect} runs, as Figure~\ref{fig:mprot2}a shows. \revised{This suggests that the performance of virtual memory (VM) operations is undesirably held back by the number of threads, even in the idealized scale-out scenario wherein these threads do not exhibit any data sharing and perform no synchronization operations such as barriers or locks.} As data centers scale beyond multi-socket systems to multi-node systems where a single process can span over multiple compute nodes connected via remote memory protocols like CXL~\cite{li2023pond, maruf2023tpp}, ensuring TLB coherence across the logical process becomes increasingly expensive. \revised{Moreover, as VM abstraction continues to play an ever more important role in simplifying programming models for emerging systems such as heterogeneous and disaggregated memory architectures~\cite{Devirtualizing,lightweightvm,infiniswap,legoos}, keeping the overhead of VM operations low is crucial to guarantee future performance.}

\revised{Apart from being over-conservative on TLB coherence operations upon page-table updates,} page-table reads, which happen during page walks as a result of TLB misses, are also significantly affected by the NUMA architecture. 
%While the relationship between the placement of data pages and the scheduling of threads that use them on NUMA sockets has been studied extensively ~\cite{calciu2017blackbox,dashti2013traffic,gandhi2014efficientvirtualization,kaestle2015shoal}, the placement of key OS data structures such as page-tables has been largely neglected until recently.
Mitosis~\cite{mitosis2019} demonstrated that the performance penalty of the requested page-table entry (PTE) being allocated on a remote NUMA node is often higher than the penalty of the requested data page being remote; this is particularly true for big data applications that experience high TLB miss rates~\cite{lightweightvm,redundantvm}.

In this work, we introduce a novel page-table management mechanism, called \sysname, that seeks to simultaneously improve the performance of both page-table \textit{READ} and \textit{UPDATE} operations in NUMA systems. \sysname enables transparent, partial, and on-demand page-table replication across NUMA nodes to ensure that address translations for local data is satisfied within the same NUMA node.
%reaping all the benefits of page-table replication.
\sysname achieves this by creating replicas of individual PTEs on the NUMA node that requests them. In doing so, \sysname avoids costly and ineffective system-wide page-table replication along with any coherence actions that would arise from such replication, as it limits the scope of replication and the related coherence actions only to nodes that actually share the same PTEs.
%As a result, \sysname significantly reduces the memory overheads of page-table replicas and avoids the scalability issues of eager replication.

We further observe that \sysname's precise and coherently maintained knowledge of which sockets contain copies of any individual page-table can be leveraged to reduce the scope of TLB shootdowns sent upon any change to that page-table \revised{--- an effective solution to the scalability bottleneck due to sending IPIs indiscriminately to all cores running the same process}. Namely, \sysname's lazy on-demand replication of page-tables ensures that the following invariant holds by design: if there is no replica of a given page-table on a given socket, there can be no thread running on that socket that currently has the corresponding PTE in its TLB; if any core on the socket has the PTE in its TLB, then the PTE must also exist in a local replica by design. As a consequence of this invariant, expensive TLB shootdowns do not need to be sent to any thread running on a socket that does not have a copy of the corresponding page-table. This allows \sysname to safely filter out many TLB shootdowns, dramatically improving the performance of memory management operations that change page-tables, as well as improving the performance of threads that avoid receiving unnecessary TLB shootdowns.

We implement \sysname in Linux (v4.17 and v6.57) and show that \sysname achieves the full benefits of page-table replication for page-table READs, as well as \revised{a 12\% and 36\% improvement in runtime on Apache Webserver and Memcached} due to more efficient page-table UPDATES. \sysname minimizes the memory footprint and page-table coherence overheads, and avoids the scalability limitations of eager replication. As shown in Figure~\ref{fig:mprot1}, \sysname is able to entirely eliminate the NUMA effect of operations such as \texttt{mprotect}, whose performance is improved by nearly 40x. We plan to open-source \sysname to encourage more research in this area.

The rest of the paper is organized as follows. Section~\ref{sec:background} presents the background on page-table management on NUMA systems. Section~\ref{sec:design} describes the design principles of \sysname. Section~\ref{sec:evaluation} presents our evaluation methodology and results. Section~\ref{sec:future} provides discussion and directions for future work. Section~\ref{sec:related} discusses related work and Section~\ref{sec:conclusion} concludes the paper.
\section{Background}
\label{sec:background}

\begin{table}
\resizebox{\columnwidth}{!}{
\begin{tabular}{l|c|c|c|c|c|c|c}
\toprule
System &
  \begin{tabular}[c]{@{}c@{}}Translation \\ Performance\end{tabular} &
  \begin{tabular}[c]{@{}c@{}}Selective \\ Replication\end{tabular} &
  \begin{tabular}[c]{@{}c@{}}Implicit \\ Policy\end{tabular} &
  \begin{tabular}[c]{@{}c@{}}Lazy \\ Replication\end{tabular} &
  \begin{tabular}[c]{@{}c@{}}Efficient \\  PTE Updates\end{tabular} &
  \begin{tabular}[c]{@{}c@{}}Efficient \\  Migration\end{tabular} &
  \begin{tabular}[c]{@{}c@{}}NUMA \\ Scalability\end{tabular} \\ 
  \midrule
Linux & {\color[HTML]{FE0000} \xmark}  & {\color[HTML]{FE0000} \xmark} & N/A   & N/A   & {\color[HTML]{FE0000} \xmark} & {\color[HTML]{FE0000} \xmark} & {\color[HTML]{FE0000} \xmark} \\ \hline
Mitosis & {\color[HTML]{32CB00}\cmark} & {\color[HTML]{FE0000} \xmark} & {\color[HTML]{FE0000} \xmark} & {\color[HTML]{FE0000} \xmark} & {\color[HTML]{FE0000} \xmark} & {\color[HTML]{32CB00}\cmark} & {\color[HTML]{FE0000} \xmark} \\ \hline
\sysname   & {\color[HTML]{32CB00}\cmark} & {\color[HTML]{32CB00}\cmark} & {\color[HTML]{32CB00}\cmark} & {\color[HTML]{32CB00}\cmark} & {\color[HTML]{32CB00}\cmark} & {\color[HTML]{32CB00}\cmark} & {\color[HTML]{32CB00}\cmark} \\ 
\bottomrule
\end{tabular}
}
\caption{Comparison of state-of-the-art solutions in support of page-tables on big NUMA machines.}
\label{table:comparison}
\end{table}

\subsection{Virtual Memory}
Page-tables are a key component of most modern operating systems and are used to map the virtual address space of a process to the physical memory available on the hardware platform. As page-tables are hierarchically organized in multiple levels, conducting a full page-table walk usually requires multiple memory accesses~\cite{denning1970virtual} and on most systems it is done in hardware for performance reasons.

Translation Lookaside Buffers (TLBs) are used to accelerate the process of address translation by caching virtual-to-physical mappings that are frequently used. Unfortunately, the growth of main memory capacity has far outpaced the growth of TLB sizes in recent years. As a result, TLB coverage has stagnated which results in a higher TLB miss rate ~\cite{basu2013efficient,karakostas2015redundant,pham2014increasing,pham2012colt,cox:effaddr,du:superpage}. When a TLB miss occurs, the Memory Management Unit (MMU) has to perform a page-table walk to retrieve the appropriate Page-Table Entry (PTE). This is a time-consuming process that can take several hundreds of cycles to complete~\cite{ryoo2017rethinkingtlb}. As a result, it is not unusual to see applications spending anywhere from 10-93\% of CPU cycles servicing TLB misses~\cite{basu2013efficient,gandhi2014efficientvirtualization,karakostas2015redundant,margaritov2019asap}, especially on NUMA systems~\cite{mitosis2019, vmitosis2021}. In addition to TLBs, modern CPUs also use Page Walk Caches (PWCs) to cache intermediate nodes of the page-tables that are frequently accessed. This allows the MMU to bypass some of the upper levels of the page-table. However, the number of entries in such caches is normally limited to several dozen entries due to hardware constraints which limits their effectiveness~\cite{yaniv2016hash, oskin:tlb}. 

\begin{figure}[t]
        \includegraphics[width=1\columnwidth]{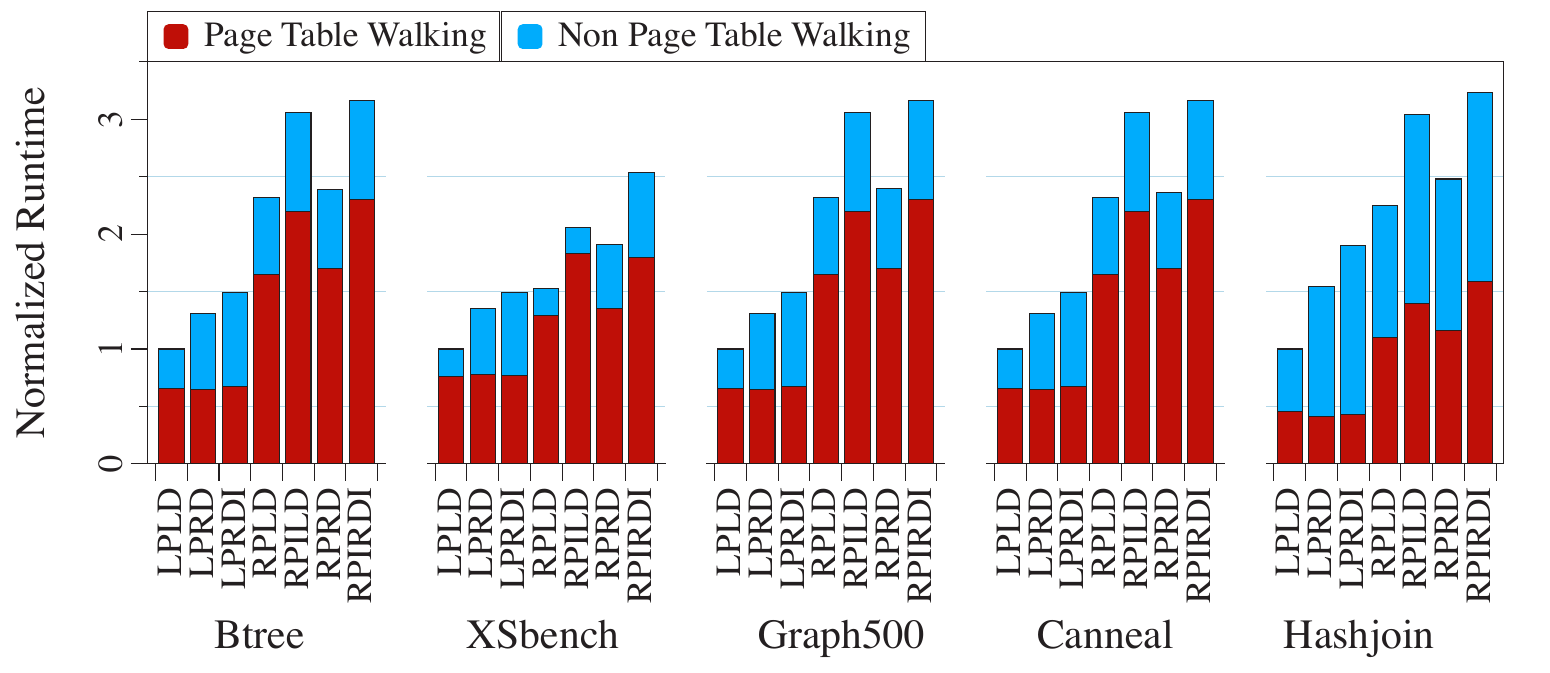}
        \caption{Impact of data and page-table placement on performance of various applications; L - local, R - remote, P - page-tables, D - data, I - interference of other applications on inter-socket traffic. The impact of page-table walks on the run time is significantly high often higher than data access. Detailed settings are show in Table~\ref{tab:workload-config}.}
        \label{fig:remoteLocal}
\end{figure}

\subsection{NUMA Architectures}
NUMA architectures have historically come in many forms, but their defining feature is that accessing memory attached to a local node will have higher bandwidth and lower latency and energy compared to accessing memory on a remote node. NUMA systems are particularly popular in data centers and cloud systems in the form of multi-sockets, as they scale well with larger memory capacity. The NUMA paradigm is today further driven by the emerging architectures based on chiplets and multi-chip modules~\cite{demir:galaxy, iyer:heterogeneous, kannan:enabling, yin:modular, amd, intel, marvell, tsmc}.

 Because of the large latency, bandwidth, and energy gap between accesses to local and remote NUMA nodes, the performance of NUMA systems largely depends on the ability of the system to maximize the chances that a piece of data is located in the same NUMA node as the thread requests it~\cite{calciu2017blackbox,dashti2013traffic,gandhi2014efficientvirtualization,kaestle2015shoal}, with most modern operating systems providing some form of support for optimizing data page placement on NUMA systems. For example, Linux employs AutoNUMA which migrates data pages to sockets that are closer to the threads that access them. Linux also provides first-touch and interleaved allocation policies which affect the initial placement of data. Under the first-touch policy, memory is allocated on the first node to access it while an interleaved policy alternates memory allocation across a set of nodes which can be defined by the user.

\subsection{Eager Page-table Replication}

Modern operating systems such as Linux offer little control over the page-table placement and mostly allocate page-tables and directories on the NUMA node that touched the corresponding data first~\cite{mitosis2019}, which is a policy known as \emph{first touch}~\cite{Lameter:NUMA}. When any other NUMA node accesses the same or neighboring data, the resulting page-table traversal will lead to expensive cross-socket memory accesses, regardless of the location of the data. The hierarchical page-table organization with 4-5 levels  present in most commercial architectures further exacerbates the problem; in the worst case, during a page walk, the requested entries at every page-table level could be allocated on a different NUMA node. 

% To effectively mitigate the issues with remote page-table reads, Mitosis replicates the entire page-table hierarchy on every socket for the entire duration of process execution, and complete page-table copies are maintained coherent across all sockets~\cite{mitosis2019}. While full replication of the page-tables on every socket is highly effective at avoiding remote memory accesses for the purpose of translation, it also brings in a number of challenges:

The impact of page-table placement on the performance of NUMA systems was first studied in Mitosis~\cite{mitosis2019}. Their work showed that page-table placement can have a significant impact on how well a NUMA system performs and can even have a bigger impact than the placement of data pages in some workloads. We have successfully reproduced their results on a larger, 8-socket machine by running the same experiments using the provided scripts. The results can be seen in Figure~\ref{fig:remoteLocal}, where we can see that page-tables being remote (RP) degrades performance almost as much as data being remote (RD), and the interference of other applications on the inter-socket traffic further increases the overhead of remote accesses dramatically.

To mitigate the overheads caused by poor placement of page-tables, Mitosis proposes the use of eager page-table replication where the page-table of a process is fully replicated across all NUMA nodes. While this method may work well for applications where almost all data is shared between all threads, it could also result in unnecessary overheads due to the need to maintain page-table coherence across all replicas even if a portion of the address space is not shared between threads. Furthermore, while the memory footprint of page-tables is usually negligible relative to the size of the data allocated (\textasciitilde0.2\%), our experiments show that the memory footprint due to the additional replicas is usually around 1.5\% of the data addressed which could translate to several gigabytes of additional memory overhead in some workloads and grows linearly with the number of sockets. 

On the other hand, \sysname is able to achieve high scalability using selective and lazy replication, where it only replicates page-tables on the NUMA nodes that the accessing threads live on. Using this approach, \sysname does not require any explicit policies from the provider/developers, whose maintenance can be cumbersome in multi-tenant scenarios. Table~\ref{table:comparison} summarizes the key differences between \sysname compared Mitosis and the baseline Linux.

\section{\sysname: Design Principles}
\label{sec:design}

In this section, we discuss the high-level goals and principles we followed and design decisions we made while designing \sysname. 

\begin{table}[ht]
\centering
\footnotesize
\begin{tabular}{l|c|c|c|c}

\toprule
Config  & CPU               & Data & Interference

& Kernel    \\
\midrule
LP-LD    & 0                                     & 0                        & Invalid               & Linux     \\
LP-RD    & 0                                     & 1                        & Invalid               & Linux     \\
LP-RDI   & 0                                     & 1                        & 1 & Linux     \\
RP-LD    & 1                                     & 1                        & Invalid               & Linux     \\
RPI-LD   & 1                                     & 1                        & 0 & Linux     \\
RP-RD    & 1                                     & 0                        &         Invalid              & Linux     \\
RPI-RDI  & 1                                     & 0                        & 0,1                   & Linux     \\
RPI-LD-M  & 1                                     & 1                        & 0 & Mitosis   \\
RPI-LD-N  & 0 $\Rightarrow$ 1 & 1                        & 0 & \sysname     \\
RPI-LD-NP & 0 $\Rightarrow$ 1 & 1                        & 0 & \sysname-Prefetching \\
\bottomrule
\end{tabular}
\caption{Configuration settings for benchmarks. L, R denote two different NUMA nodes in the system (e.g., L: socket 0, R: socket 1). Note that page-table is fixed on socket 0 if no replication involved. H: \sysname, M: Mitosis, HP: \sysname with Prefetching.}.
\label{tab:workload-config}
\end{table}

\subsection{Replication Policies}
Eager replication of complete page-tables on all NUMA sockets, as done in Mitosis, is highly effective at avoiding any remote page-table accesses. This style of replication is also the least complex to implement. However, maintaining complete page-table trees in all sockets and keeping them coherent results in memory and coherence overheads that grow with the number of sockets. 

\begin{figure}
        \includegraphics[width=1\columnwidth]{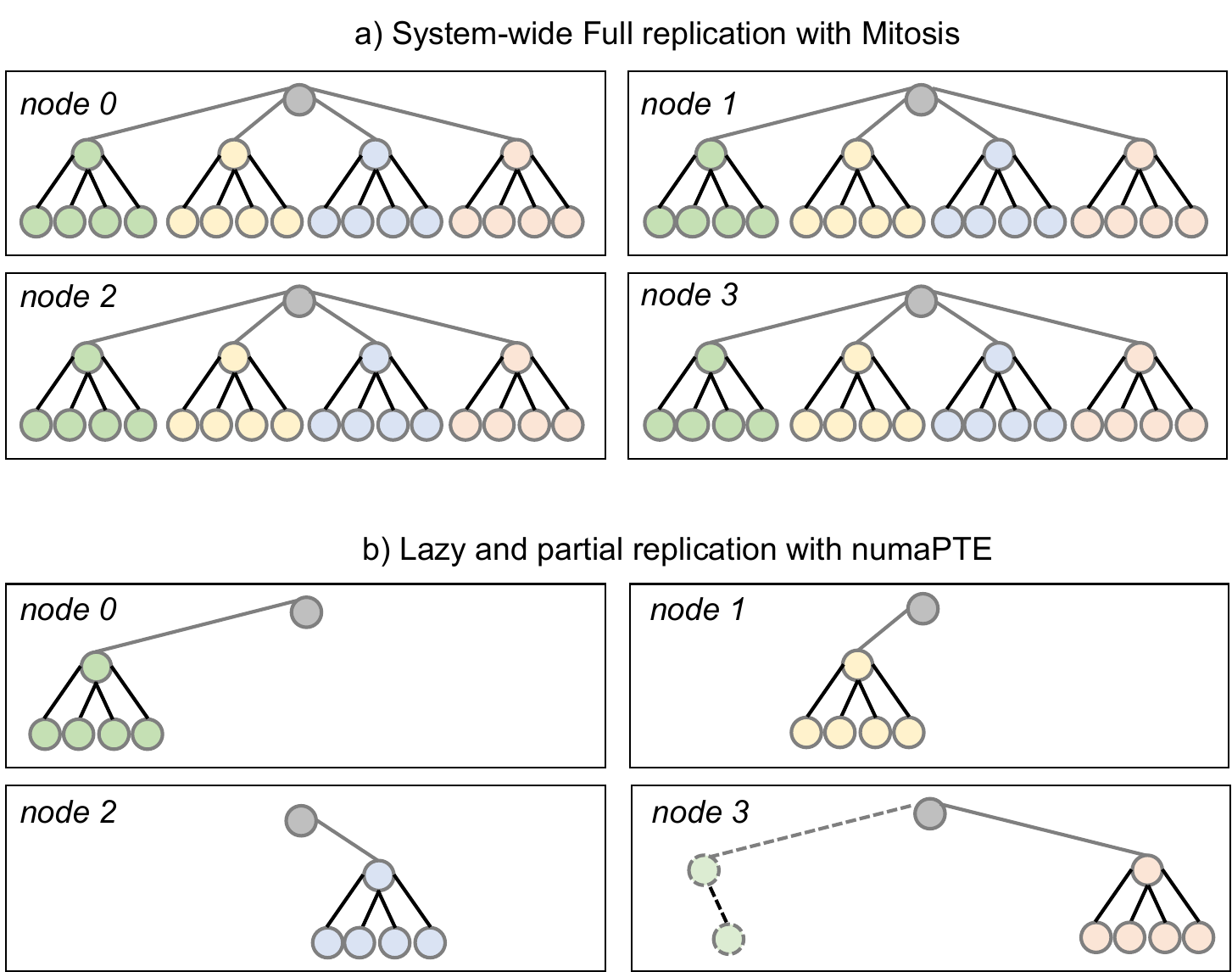}
        \caption{An abstract Illustration of replication of hierarchical page-tables on Mitosis (a) and \sysname (b). Different memory allocations (VMAs) are colored differently. Mitosis eagerly replicates allocated page-tables on all NUMA nodes (sockets), whereas \sysname performs \textit{lazy and partial} replication on-demand simultaneously instead.} 
        \label{fig:policies}
\end{figure}

To reduce these overheads in the future, Mitosis envisions the use of \emph{explicit} page-table allocation policies, by which the user would be able to specify a subset of the NUMA nodes on which the page-tables will be fully replicated; the other NUMA nodes would not have any page-tables of that process. We call this style of replication \emph{selective} replication, where the user (or the system) can select the NUMA nodes they want to limit the replication. We argue against such policies, because they put the burden on the user and/or the system to ensure that both the execution threads and the corresponding page-tables are co-located on the same NUMA node throughout the execution of the process. In case of any deviation, e.g., when some threads migrate, the system would need to decide if, how, and when to establish a complete new copy of the page-tables on a new node, to support local address translation on the new socket. The system would also need to decide how and at which point to shut down a full copy of page-tables on a socket that a number of threads (or all threads) migrated from, in order to reduce memory and coherence overheads. While these policies could be implementable and would reduce some of the Mitosis overhead, they put a lot of burden on the user and/or the system to follow them and promptly adjust to any deviations. Although these policies could reduce the overheads compared to full-system replication by reducing the number of replicas, each replica still remains \emph{complete}, resulting in memory and coherence overheads that eager replication cannot reduce any further.

\sysname, in contrast, uses lazy and \emph{partial} replication, by which only the individual PTE entries that are demanded are allocated and copied to a new node. This automatically ensures 1) the minimum amount of memory spent on replication, 2) the minimum amount of activity needed to keep the replicas coherent, 3) the minimum overhead of aggregating dirty/accessed bits written by hardware in different PTE replicas, which happens when these bits are read by the OS. Importantly, our lazy approach that piggybacks on the standard page-fault procedure automatically ensures that PTEs are replicated wherever the thread is executed; when the thread is migrated the new PTE copies will be automatically established. In other words, \sysname enables selective (only on relevant nodes) and partial (only the relevant PTEs) replication in a lazy manner that automatically guarantees the co-location of execution threads and page-tables, sidestepping the problems of eager replication and explicit policies.

\subsection{Page-Table Coherence Protocol}
The challenge with lazy and partial page-table replication is that at the time when the requested PTE is missing, the OS needs to find out which node, if any, has a copy of the target PTE. Therefore, a mechanism that finds at least one existing sharer is needed, and if such a sharer does not exist, \sysname needs to know that. 

One solution would be to statically designate one node to hold an entire copy of all page-tables, in which case partial replicas can be built lazily on the other nodes by copying the PTEs from the designated node on demand. However, the main problem with such a solution is that every page-table used has an extra sharer (the master node) that generally does not need that page-table for the translation of its own data, but nonetheless must be kept up to date. Other problems include load imbalance and the fact that the traffic to the destination node can easily become congested, either because of too much translation-related traffic or due to interference of other applications, slowing down the whole system. Furthermore, a single source node becomes a single point of failure and presents the scalability bottleneck. 
 
We instead propose an efficient and decentralized page-table coherence protocol in which every memory allocation (e.g., virtual memory area or VMA in Linux) is assigned an \emph{owner}; the owner of each allocation area is the NUMA socket that requested its allocation. We maintain the following invariant: if a valid PTE for a given page exists, the owner node must have it. This invariant is needed because a circular list of sharers is efficiently maintained at the level of individual page-tables~\cite{mitosis2019}, and we would not be able to reach that list of sharers if we do not know at least one node that is currently in the list. When the requested PTE does not exist on the local node, the entry is copied from the \emph{owner} node as part of the page-fault handler, and the new entry is added to the list of sharers of that page-table. If the owner does not have that entry, then it means that the page has not yet been touched, so the PTE is created by taking a page fault on that page, and the entry is then inserted into both the owner's and replica's copy of the page-table and the two copies of the page-tables are linked in a circular list, if they were not linked before (for example, due to other PTEs in the same page-table).

In the case of system-wide full page-table replication~\cite{mitosis2019}, any changes to page-tables must be propagated to all replicas on every socket in the system. Because each replica is located on a different socket, updating all of them can take a significant amount of time. Furthermore, these updates must be performed while holding already contended memory-management locks, significantly affecting other page-table management operations happening concurrently, as we will show in Section~\ref{sec:evaluation}. In the case of \sysname, upon any change to a page-table, only the replicas found in the list of sharers for that particular page-table are updated, if any. The coherence actions are therefore limited only to page-tables that are actually shared, and the cost of the actions is limited by the number of nodes that actually share the page-table.

\subsection{Replication and Partitioning}
As an illustrative example, assume that a process with four threads runs on a 4-socket NUMA machine, one thread per socket. Also, assume that each thread is allocating a chunk of memory in its local memory and accesses only the data it allocated. This is an ideal scenario from a NUMA system point of view, as it enables perfect data parallelism. Figure~\ref{fig:policies}a illustrates a potential state of page-tables in such a scenario in case of eager system-wide page-table replication, while Figure~\ref{fig:policies}b shows the state of page-tables in case of \sysname. For simplicity of illustration, the page-table is depicted as a radix-tree with three levels of the hierarchy and an out-degree of four (in practice, there are 4-5 levels with an out-degree of 512). As shown in Figure~\ref{fig:policies}a, full and eager replication will replicate all parts of the page-tables on all sockets. On the other hand, \sysname's protocol assigns ownership based on the allocation to the node that allocated the VMA, and as a result, each node is by default the owner of its own page-table replica. Unless a thread from one node tries to access data allocated by a thread on a different node, there will be no page-table replication in the system except for the root node. This effectively \emph{partitions} the page-table into four independent partitions, each co-located with its data and accessed and managed locally without any cross-socket coherence activity between them, and this is all achieved automatically without intervention from the user or the system.

Apart from support for perfect data partitioning, \sysname also enables data sharing much more efficiently compared to Mitosis. Let's assume that node 3 accesses a piece of data located on node 0 for the first time. In \sysname, this will lead to the allocation of a single page-table on node 3 and the missing PTE will be copied, as shown in Figure~\ref{fig:policies}b. Any updates to this page-table would limit the coherence activity only to two nodes involved, as opposed to system-wide. 

\subsection{Configuring Laziness} %FIXME
When the requested PTE entry does not exist on the local node, \sysname copies only the requested PTE entry from the owner. This is the laziest form of \sysname that we support, as no PTE is replicated unless demanded by a NUMA node. However, one could expect that such extreme laziness has a price, as the very first access to a page from a new node is guaranteed to result in a remote page-table access; this is in contrast to Mitosis, which eagerly replicates every page-table entry before it gets to be used for the first time and always ensures local page-table access.

Interestingly, we have found that extreme laziness incurs virtually no penalty in our workloads. The simple reason is that \sysname's laziness is penalized only upon the very first access to a given page by a new socket; any subsequent accesses to any part of that page by any core running on the same socket will find that PTE in the local memory. As the average number of times a page is accessed during its residence in memory is significantly higher than one for all the workloads we experimented with, we conclude that the performance penalty for laziness is negligible for most applications. 

\begin{figure}
        \includegraphics[width=1\columnwidth]{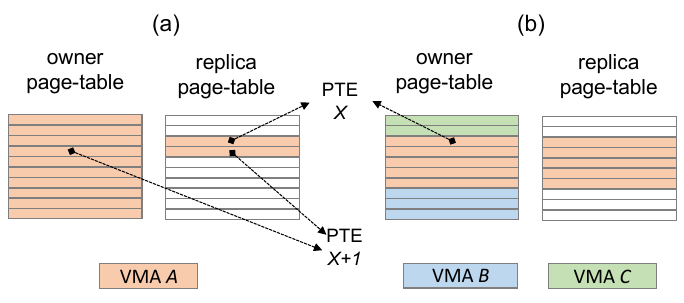}
        \caption{(a) Prefetching with a degree of 1 with PTE $X+1$ prefetched. (b) Maximum degree prefetching when the target page-table covers multiple VMAs is limited by both the page-table boundaries and boundaries of the encompassing VMA.}
        \label{fig:prefetch}
\end{figure}

However, there are applications with low temporal reuse of pages that could benefit from a less lazy approach (we construct such a worst-case microbenchmark in Section~\ref{sec:evaluation}). Furthermore, as we will show in Section~\ref{sec:evaluation}, extreme sharing of data where every page ends up being shared on every socket, and every page-table is correspondingly allocated on every socket could effectively turn \sysname into Mitosis, with full replicas present on every socket. To improve \sysname's performance in such scenarios and reduce its laziness, we provide a low-cost support for prefetching a configurable number neighboring page-table entries together with a requested PTE, to maximize the chance of a local page-table access upon the very first access to that page by a new node. In our implementation, we provide an additional kernel parameter that allows the user to indicate the desired degree $d$ of prefetching. Prefetching degree of $d$ will fetch $2^{d}$ neighboring entries including the desired PTE (for $d$=0, in total $2^{0}$=1 entry will be fetched, i.e., the requested PTE only). The prefetching is limited to the full page-table, as logically consecutive page-tables in the virtual address space are not necessarily physically adjacent. We also limit prefetching to the VMA boundaries, as other neighboring VMAs are not necessarily logically related, which is illustrated in Figure~\ref{fig:prefetch}b.

\subsubsection{Prefetching Overheads}
The overhead of physically copying additional consecutive PTEs from the same page-table is negligible and we could not observe it in our measurements. At the same time, PTE prefetching also does not cause any PTE coherence overheads; to understand why, let's look at Figure~\ref{fig:prefetch}a, which shows an example of one page-table that covers a fraction of single virtual memory area VMA $A$. Assume that the replica requests PTE $X$ that exists in the owner node but not in the replica. Also assume that the replica does not have this exact page-table allocated yet. Upon a page-fault on PTE $X$ on the replica node, \sysname would by default fetch only PTE $X$ from the owner, but in this example we choose a degree of  $d$=1, with a total of $2^{1}$=2 PTEs to be fetched, one of which, PTE $X+1$, is prefetched. Assume that PTE $X+1$ ends up never being used by the replica. The question is whether fetching PTE $X+1$ causes any additional coherence action that otherwise would not have happened. 

 Recall that \sysname does not keep the list of sharers at the level of individual PTE entries, which would be impractical and unnecessary, but at the level of individual page-tables, similar to Mitosis~\cite{mitosis2019, vmitosis2021}. As a result, when PTE $X+1$ is modified by the owner, the owner node cannot know if the replica has PTE $X+1$ or not; it only knows that the replica has $some$ PTE entries from the page-table in question. Therefore, the replica must be updated regardless of whether or not PTE $X+1$ exists on the replica, and prefetching PTE $X+1$ does not cause any additional coherence actions. Given the negligible overheads of prefetching and potentially sizeable benefits in some applications, we suggest to \sysname users to set the degree of prefetching to maximum by default.

\subsection{Reducing TLB shootdowns}
When page-table entries are changed for any reason (e.g., due to the \texttt{mprotect} system call, \texttt{munmap}, a page is swapped out, etc.), the corresponding entries in the TLBs of each core that currently runs a thread of the same process have to be invalidated to ensure that stale cached page-table entries are not used in translation. This is done by sending an inter-processor interrupt (IPI) to each core that is running a thread from the same process, which causes significant overhead for all involved cores, particularly so for the initiating core. This process is known as a \emph{TLB shootdown}. TLB shootdowns are known to be very expensive in general~\cite{kumar:latr, oskin:tlb} and particularly so for NUMA systems~\cite{amit:optimizing}, where they pose a scalability challenge.

The reason that TLB shootdowns must be sent to every thread of the same process is that the OS cannot know which TLBs contain any given entry. However, \sysname has precise information about what which NUMA nodes contain a copy of any given page-table. Also note that, by design, it is not possible for any NUMA node to contain a given PTE entry in any of its TLBs unless the NUMA node is in the list of sharers for the page-table encompassing that PTE (if any TLB contained the PTE in question, then that TLB entry must be have been filled from a local copy, or the local copy would be filled together with the TLB). Therefore, \sysname can use the sharer information to safely reduce the scope of the TLB shootdowns and not issue them to any cores on those NUMA nodes that are not in the list of sharers for the particular page-table, because these nodes are guaranteed by design not to have the PTE in any of their TLBs.

\section{Evaluation}
\label{sec:evaluation}

% machine detail
\subsection{Evaluation Platform}
\label{subsec:hwspec}
We conducted all measurements on an eight-socket NUMA machine with 8TB DDR4 physical memory in total. Every socket is equipped with 1TB DDR4 memory and one Intel Xeon E7-8890 v3 processor with 18 cores, each of which works at a base frequency of 2.5 GHz and has two hyper-threads. Each processor has a unified 45MB L3 cache, a unified 256 KB L2 cache, a unified L2 TLB with 1024 entries and a private L1 TLB with 64 entries for each core. Hyper-threading is enabled and turbo-boost is disabled in all our experiments. 

\begin{table}
\footnotesize
\begin{tabular}{ lp{0.35\textwidth} lp{0.3\textwidth} }

% \hline
\toprule
\textbf{Workload} & \textbf{Description}  \\ 
\hline
XSBench \cite{tramm2014xsbench} & Monte Carlo neutron transport computational kernel applications. Dataset = 85GB, p=50M, g=200k. Only used in multi sockets experiment.\\ % Content row 2
\hline

Graph500~\cite{murphy2010introducing} &  Benchmark for generation, compression and search of large graphs. Dataset = 160GB, s=30, e=20.\\ % Content row 4
\hline
Redis \cite{redis} & Single thread in-memory data structure store. Dataset = 256GB, key size = 25, element size = 64, element number = 1B, 100\% reads. Only use in workload migration.\\ 
\hline
Btree~\cite{mitosis2019} & Benchmarks for measuring the index lookup performance in large applications. Dataset = 110GB, 1M keys, 10B lookups. \\ % Content row 7
\hline
HashJoin \cite{chen2007improving} & Benchmark for hash-table probing in database. Dataset = 145GB, 10B elements. \\ % Content row 8
\hline
Canneal~\cite{bienia2009parsec}  & Simulates routing cost optimization in chip design. Dataset = 110GB, 400M elements. \\
% \hline
\bottomrule % Bottom horizontal line
\end{tabular}
%\end{adjustbox}
\caption{Detailed description of the workloads.}
\label{tab:workloads} 
\end{table}

\begin{table}[]
\resizebox{\columnwidth}{!}{
\begin{tabular}{l|l|c|c|c|c|c} 
\toprule
\multicolumn{2}{l|}{Workloads}                                                     & Graph500 & Btree & HashJoin & XSBench & Canneal  \\ 
\hline
\multicolumn{2}{l|}{\begin{tabular}[c]{@{}l@{}}Program Size (GB)\end{tabular}}   & 160      & 110   & 145      & 85      & 110      \\ 
\hline
\multirow{3}{*}{page-table footprint (GB)} & Linux                                      & 0.31     & 0.22  & 0.28     & 0.17    & 0.22     \\ 
\cline{2-7}
                                      & Mitosis                                    & 2.51     & 1.72  & 2.27     & 1.33    & 1.72     \\ 
\cline{2-7}
                                      & \sysname                                      & 0.67     & 0.44  & 0.4      & 1.32    & 0.32     \\ 
\hline
\multirow{3}{*}{page-table overhead (\%)}  & Linux                                      & 0.2      & 0.2   & 0.2      & 0.2     & 0.2      \\ 
\cline{2-7}
                                      & Mitosis                                    & 1.56     & 1.57  & 1.57     & 1.57    & 1.57     \\ 
\cline{2-7}
                                      & \sysname                                      & 0.42     & 0.4   & 0.28     & 1.55    & 0.29     \\
\bottomrule
\end{tabular}} \caption{Page-table footprint (GB) in the baseline (no replication, Mitosis~\cite{mitosis2019}, and \sysname for various benchmarks.}
\label{tab:ptsize} 
\end{table}

\subsection{Page-table Footprint}
\label{sec:ptf}
Table~\ref{tab:ptsize} shows the program size and total page-table footprint for various workloads running on an 8-socket machine for three configurations: the baseline (no replication), Mitosis, and \sysname. As expected, Mitosis consistently results in 8x larger page-table footprint compared to the baseline (7x overhead). In contrast, the page-table footprint overhead in \sysname is only between 0.5x-1x and corresponds to the level of actual data sharing across sockets. The only exception is XSbench, which has extreme data sharing and every page-table is replicated on every socket, producing the same page-table footprint in the case of \sysname and Mitosis. In this case, \sysname effectively converges to Mitosis. 

Note that \sysname's PTE prefetching has no impact whatsoever on page-table memory footprint, because the prefetching is limited to the page boundaries surrounding the requested PTE. At the time of prefetching the replica page for the requested PTE has already been allocated and the whole page is accounted for in the footprint, regardless of prefetching.

\subsection{PTE Prefetching}

\begin{figure}
        \includegraphics[width=1\columnwidth]{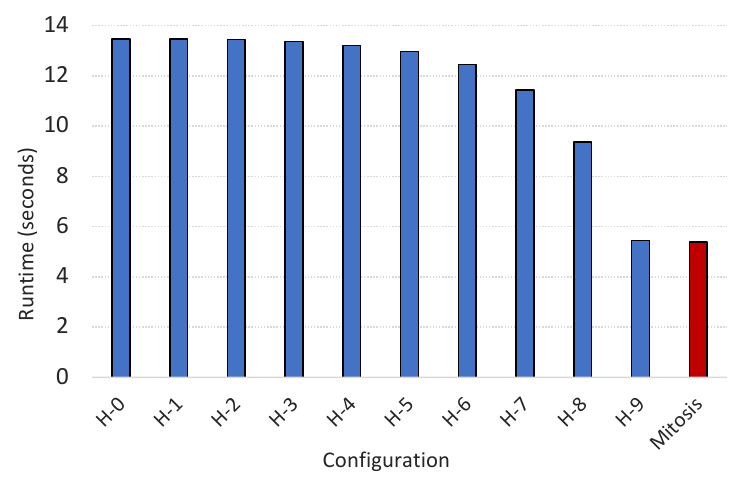}
        \vspace{-0.3in}
        \caption{The time taken to traverse all pages in a large segment exactly once for various configurations. H-$i$ refers to \sysname with a degree of prefetching $i$. } 
        \label{fig:prefetching2}
\end{figure}

To demonstrate the potential of PTE prefetching, we construct a microbenchmark that traverses, in a random order, a 1GB array such that every page is accessed exactly once, which is the worst case for \sysname. The array is set up and initialized on one node, and then it is accessed on the other. The benchmark is constructed such that it achieves a near-zero hit ratio in caches and TLBs, and therefore nearly every data access results in a remote memory access for the purpose of translation.

Figure~\ref{fig:prefetching2} shows the time it takes to traverse the array in the random order in the case of \sysname with multiple levels of PTE prefetching (from 0 - no prefetching, to 9 - maximum prefetching of $2^{9}$=512 PTEs. We also show the results for Mitosis for comparison. We can see that prefetching within a page-table is enough to eliminate any laziness penalty that \sysname experiences by lazily copying PTEs one by one. Also note that subsequent traversals of this array would lead to identical behavior in the case of \sysname and Mitosis regardless of the level of prefetching, because at that point all page-tables are constructed and replicated where they are needed.

\subsection{Workload Migration}

\begin{figure}[t]
        \includegraphics[width=1\columnwidth]{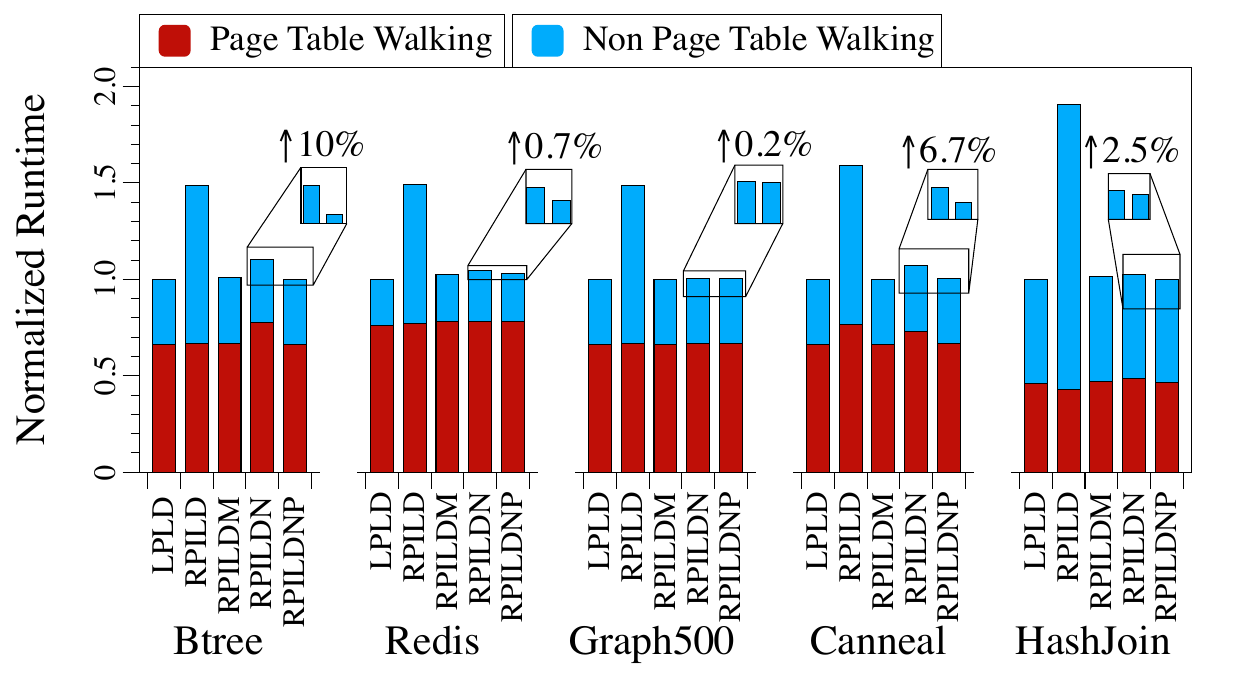}
        \caption{Normalized performance in the workload migration scenario for Mitosis, \sysname, and \sysname with a prefetching degree of 9. All of the configurations shown on X-axis are listed in Table~\ref{tab:workload-config}.} 
        \label{fig:workload-migration}
\end{figure}

Figure~\ref{fig:workload-migration} shows the behavior of Linux, Mitosis and \sysname in a scenario where a thread migrates to a different socket, where the data resides. In the case of Linux, the page-tables will remain on a remote socket, and the thread will be accessing the remote socket only for translation. This causes a significant slowdown in the presence of applications that interfere with inter-socket traffic (RPILD). However, Mitosis doesn't suffer from this problem as it pre-replicates the page-tables system-wide. \sysname suffers from a small performance penalty due to lazy replication (RPILDN), but that penalty is largely eliminated through prefetching.

\begin{figure}
        \includegraphics[width=1\columnwidth]{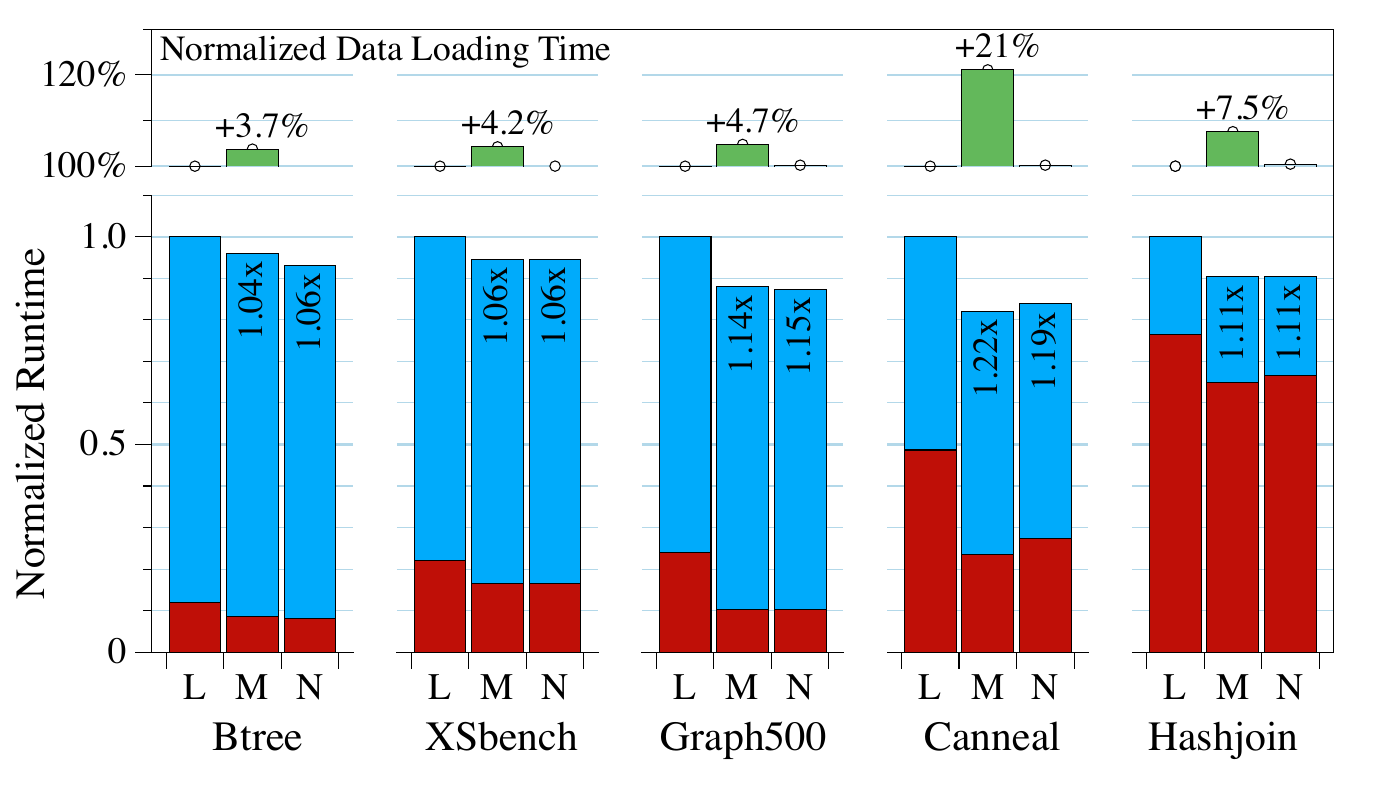}
        \caption{The lower part shows performance of Mitosis and \sysname for the workloads in a multi-socket setting, normalized to the baseline Linux. L, M and N indicate Linux, Mitosis, and \sysname, respectively. The higher part shows the data loading time in the same setting, also normalized to the base Linux.}
        \label{fig:multisocket}
\end{figure}

\subsection{General Applications}
Figure~\ref{fig:multisocket} shows the performance of multiple real-world applications, listed in Table~\ref{tab:workloads}, running on our 8-socket machine. The workloads consist of two phases: 1) data setup/loading, which exercises page-table updates, and 2) the execution phase with the loaded in-memory data, which stresses page-table reads. During the loading time (the upper plot), all page-tables are constructed, and in the case of Mitosis, replicated system-wide. The execution phase (the lower plot) exhibits a significant amount of data sharing and benefits from page-table replication across sockets; once the data is loaded, there are very few page-table modifications. Note that the data setup phase is time-consuming and takes several hours, whereas the subsequent execution is benchmarked for 5 minutes. We show three configurations: baseline (no replication), \sysname (maximum PTE prefetching enabled), and Mitosis.

From the lower part of Figure~\ref{fig:multisocket} we can see that in all benchmarks the performance of \sysname matches that of Mitosis despite its laziness. In some workloads, such as Btree and Graph500, \sysname even achieves a tiny speedup compared to Mitosis due to more efficient page-table coherence, to the extent to which these workloads exercise memory management. The biggest advantage Mitosis has over \sysname is for Canneal, where Mitosis achieves 1.22x speedup over Linux, whereas \sysname achieves 1.19x. However, this advantage seems to be coming at the expense of the data loading time, during which Mitosis creates 5x more replicas than needed, according to Table~\ref{tab:ptsize} which compares the page-table footprints of all applications. This unnecessary replication results in a 21\% slowdown compared to Linux. \sysname, on the other hand, always matches the performance of Linux when it comes to data loading time, because it does not perform any replication in that phase.

\subsection{Memory Management}

\begin{figure}
        \includegraphics[width=1\columnwidth]{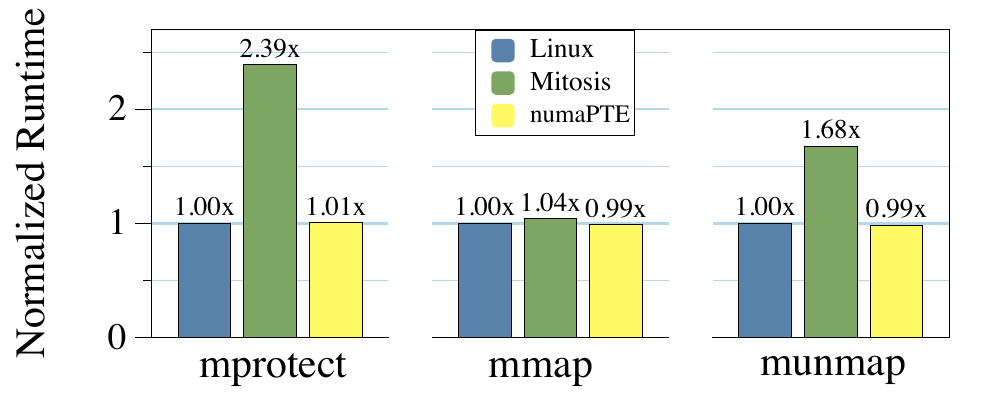}
        \caption{mprotect mmap, and munmap overhead on Linux, Mitosis and \sysname when operating on a 128K memory range.} 
        \label{fig:syscall}
\end{figure}

Figure~\ref{fig:syscall} shows the overhead of the basic memory management operations, \texttt{mmap}, \texttt{munmap}, and \texttt{mprotect} as a function of the size of the input range when executed on an 8-socket machine on three different designs: baseline (no replication), Mitosis (system-wide eager replication), and \sysname (partial and lazy replication). Note that there are no interfering threads. We see that \texttt{mmap} is largely unaffected by replication, as page-table updates are a small part of its functionality. In contrast, for mprotect and munmap, Mitosis pays a significant cost for page-table coherence, which \sysname avoids.

\subsubsection{TLB Shootdowns}
We next measure the overhead of Mitosis, \sysname without the TLB optimization, and \sysname with the TLB optimization, on the performance of the \texttt{mprotect}. The size of the \texttt{mprotect} range is a single 4KB page. The \texttt{mprotect} syscall simply flips a single bit in one PTE, and does this in a loop. Additionally, we run a varying number of spinning threads on every socket. The spinning threads have nothing to do with the \texttt{mprotect} thread; they simply increase a private counter in an infinite loop, and we measure the impact of \texttt{mprotect} on the performance of spinning threads and vice versa. The results are shown in Figure~\ref{fig:mprot1}.

When there are no spinning threads, Mitosis results in a around 25\% slowdown on 8-sockets, because it must update all replicas upon every \texttt{mprotect} operation, whereas \sysname has no overhead due to the absence of coherence activity. However, as we add spinning threads and increase their number per socket, all systems, including the baseline Linux, Mitosis, and \sysname (without the TLB optimization) will result in significant overheads, up to 40x. This is despite the fact that the spinning threads have nothing to do with \texttt{mprotect}. \sysname without TLB optimization is only slightly better than Mitosis, as it avoids updating any replicas, but still suffers a significant slowdown. This is due to the TLB shootdowns that must be sent to every running thread of the process, despite the fact that these are spinning threads. In contrast, \sysname enabled with the TLB shootdown optimization avoids sending shootdowns to any other socket, leading to zero overhead overall compared to the baseline without replication.

\begin{figure}
        \includegraphics[width=1\columnwidth]{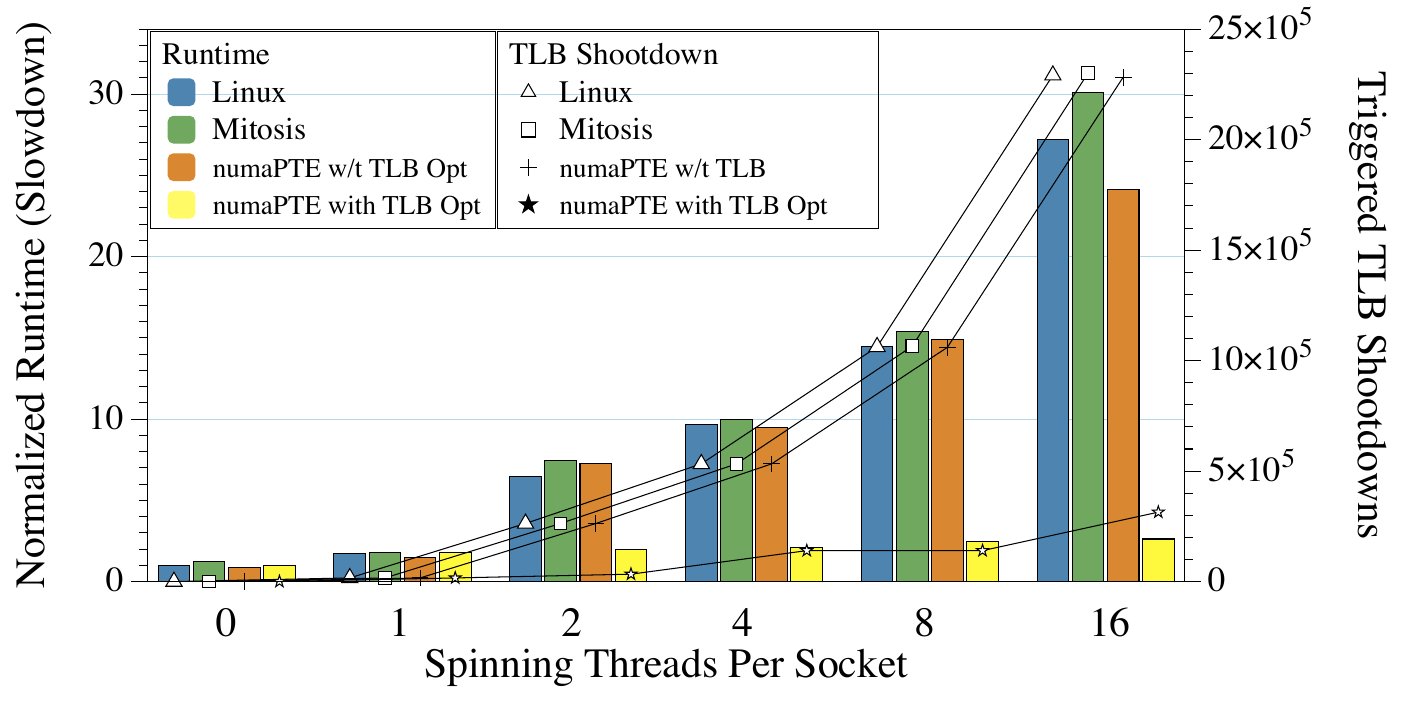}
        \caption{Impact of page-table replication and the TLB shootdown optimization on \texttt{munmap}. Existing solution are unable to quickly deallocate page-tables while \sysname does this efficiently. All values are normalized to the baseline without replication.} 
        \label{fig:unmap1}
\end{figure}

Figure~\ref{fig:unmap1} shows similar results in the same setup, except that instead of \texttt{mprotect} we use \texttt{munmap}, the underlying implementation of memory freeing. The range of \texttt{munmap} is set to a single 4KB page. With no spinning threads, \sysname does not experience any slowdowns, but Mitosis experiences a 23\% slowdown. As we add spinning threads, the overhead of Mitosis grows to almost 30x, whereas \sysname with the TLB shootdown optimization results in only 2.6x overhead. The effect of replication on \texttt{mmap} is less pronounced due to the additional work that \texttt{mmap} does that overshadows the coherence overhead.

\subsubsection{Case Study: \textbf{Memory Allocation}}

\begin{figure}[htpb]
    % \begin{minipage}[htpb]{0.33\linewidth}
        \centering
        \includegraphics[width=1\columnwidth]{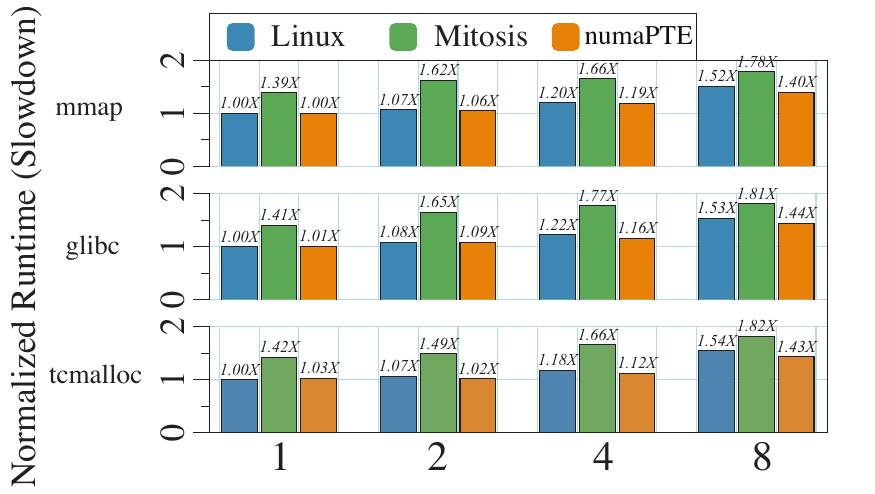}
    % \end{minipage}
    \caption{The impact of replication on $stateless$ memory allocation for various configurations. X-axis indicates the socket number.}
    \label{fig:malloc1}
\end{figure}

\begin{figure}[htpb]
    % \begin{minipage}[htpb]{0.33\linewidth}
        \centering
        \includegraphics[width=1\columnwidth]{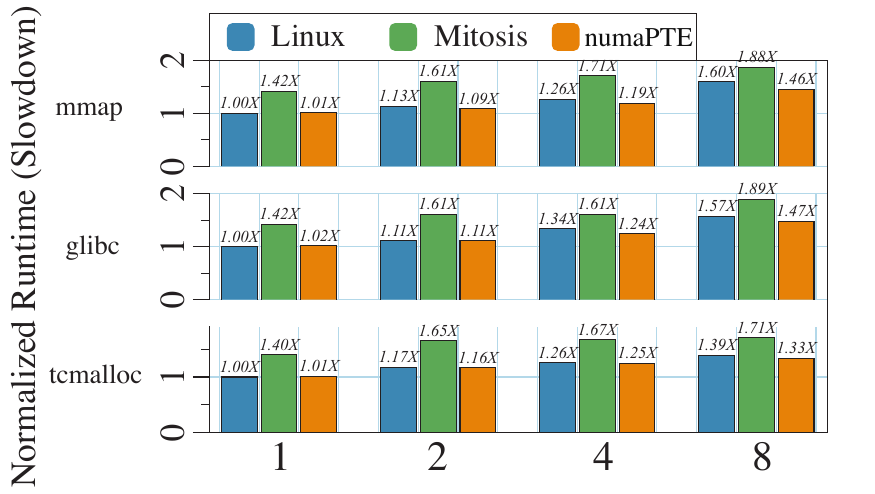}
    \caption{The impact of replication on $stateful$ memory allocation for various configurations. X-axis indicates the socket number.}
    \label{fig:malloc2}
\end{figure}

Figure~\ref{fig:malloc1} and Figure~\ref{fig:malloc2} show the overall impact of replication on memory allocation (\texttt{malloc}) on various numbers of sockets, with one thread per socket. We use three prominent \texttt{malloc} implementations: \texttt{mmap}, \texttt{glibc}, and \texttt{tcmalloc}. We develop two \texttt{malloc} benchmarks. The first benchmark is $stateless$: the benchmark will repeatedly allocate a segment of random size following the Gamma distribution with an average allocation size of about 3.3MB, and then free the allocated segment. The second benchmark follows the same allocation size distribution, but it first allocates 256 segments, and then in a loop frees one segment and allocates another, such that there are 256 concurrent allocations at any time for every thread. We can see that Mitosis results in an overhead that ranges from 1.4X to 1.9X in both malloc benchmarks. At the same time, \sysname achieves a speedup compared to both Linux and Mitosis, thanks to \sysname's minimal page-table coherence.

\subsubsection{Case Study: \textbf{Web Server}}

\begin{figure*}
    \begin{minipage}{0.5\linewidth}
        \centering
        \includegraphics[width=3in]{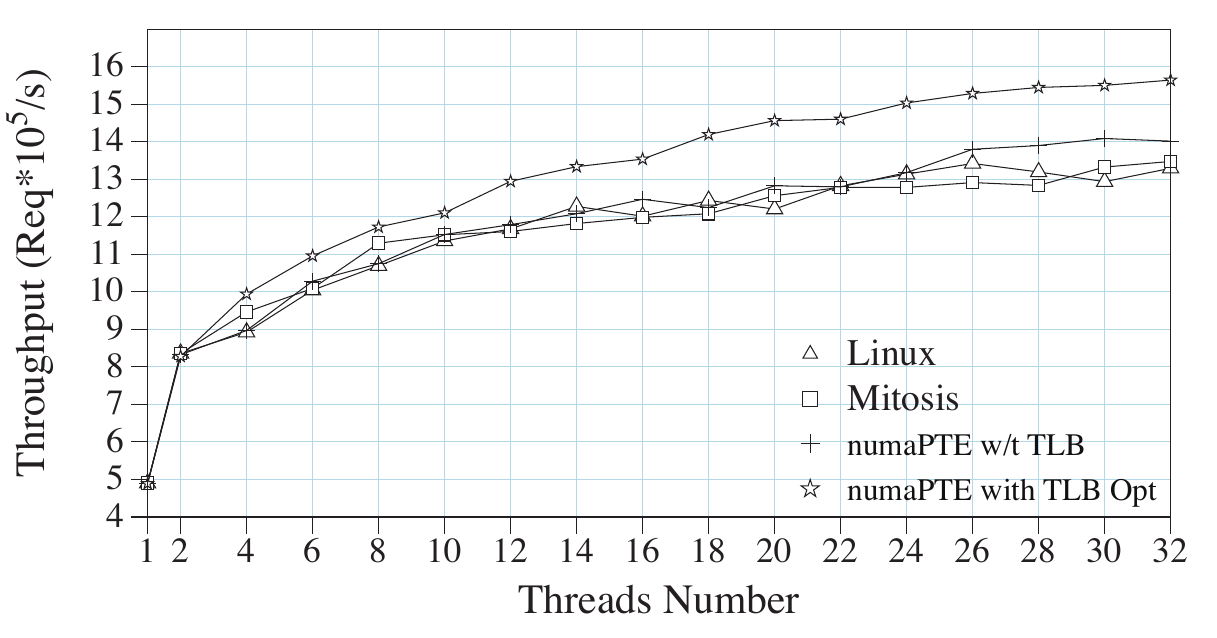} 
        \vspace{-4mm}
        \begin{center}
            a)
        \end{center}  
        \vspace{-6mm}
    \end{minipage}
\hfill
    \begin{minipage}{0.5\linewidth}
        \centering
        \includegraphics[width=3in]{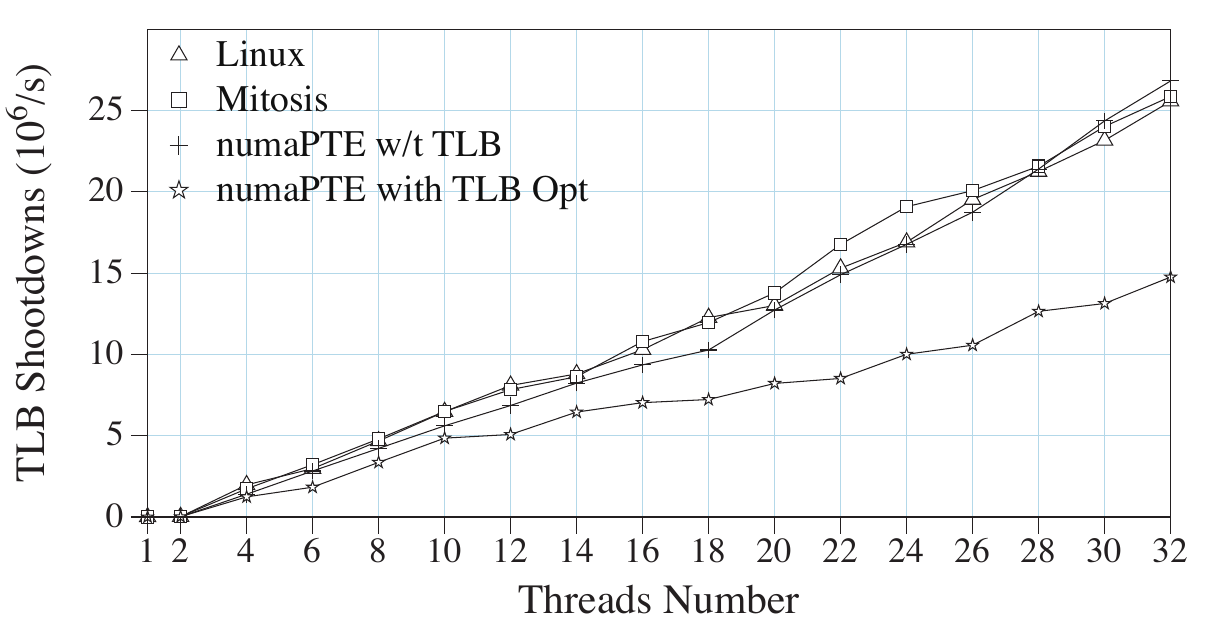}
        \vspace{-4mm}
        \begin{center}
            b)
        \end{center}  
        \vspace{-6mm}   
    \end{minipage}
\hfill 
    
    \caption{a, b: Impact of page-table replication and the TLB shootdown optimization on webserver. Normalized to the baseline without replication. The threads are evenly distributed across four sockets.
    %c: Approximate execution breakdown of webserver.
    }
    \label{fig:apache}
\end{figure*}

\begin{figure}
        \includegraphics[width=1\columnwidth]{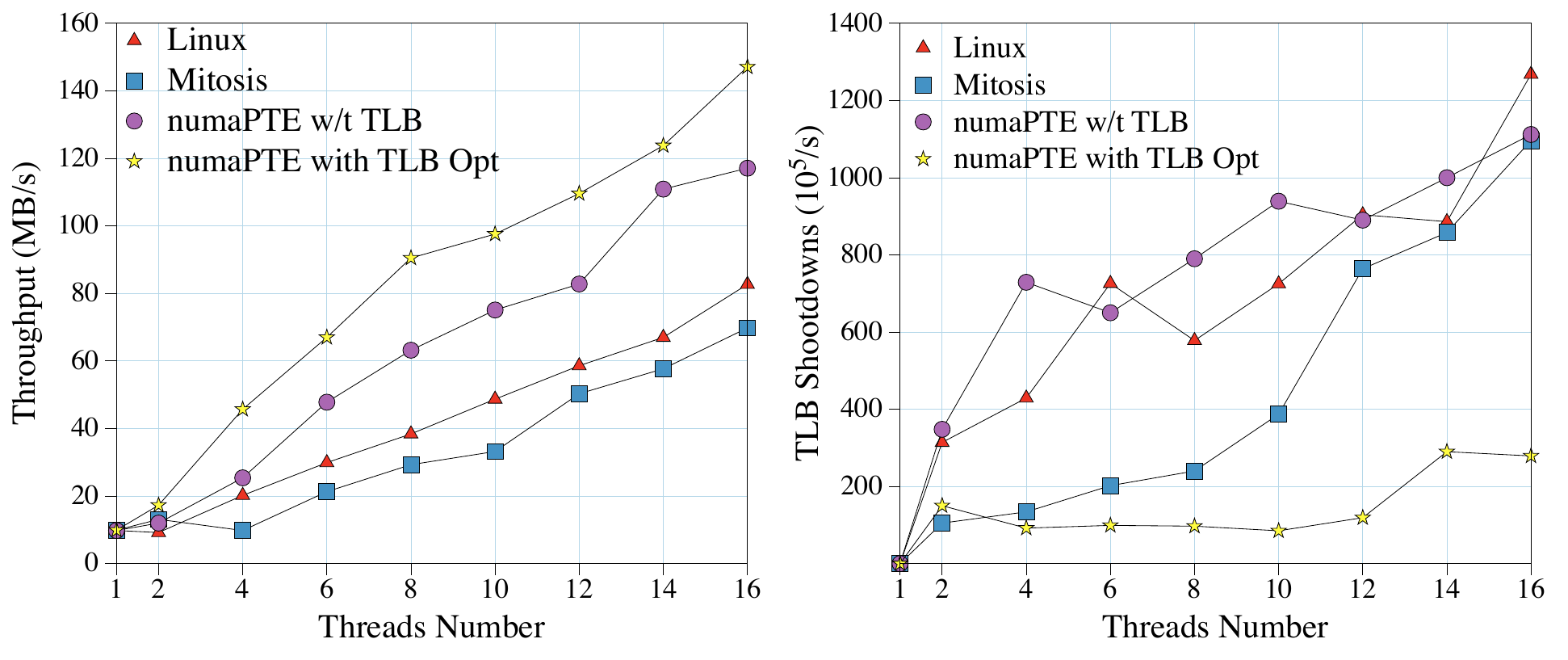}
        \caption{Impact of page-table replication and TLB shootdown optimization on Memcached running on four sockets.}
        \label{fig:memcached}
\end{figure}

Webserver, such as Apache, is an important and widely used application that is known to have problems with TLB shootdowns. Webserver application (e.g., Apache Webserver) spawns a large number of threads, each of them serving a web request, which in our case is a web page the same as in ~\cite{kumar:latr, amit:optimizing}. For each request, the webserver serving thread allocates memory for the web page using \texttt{mmap} and subsequently frees it using \texttt{munmap}, generating many unnecessary TLB shootdowns along the way. Previous work has demonstrated that techniques that reduce TLB shootdowns can significantly improve the Apache's throughput~\cite{kumar:latr}. Because of the complex NUMA impact of the NICs~\cite{neugebauer2018understanding, cai2021understanding}, Apache Webserver scales very poorly beyond a single socket, which is why constructed a synthetic benchmark that performs the webserver functionality without using a NIC, similar to prior work~\cite{neugebauer2018understanding, cai2021understanding}. 

Figure~\ref{fig:apache}a shows the throughput of the webserver benchmark  on unmodified Linux, Mitosis, \sysname without the TLB optimization, and \sysname with the TLB optimization. We run the benchmark with a varying number of threads (up to 32) uniformly distributed across four sockets. Figure~\ref{fig:apache}b shows the number of TLB shootdowns (in millions per second)  as we vary the number of threads. Because this application does not exhibit any data sharing, the impact of page-table replication on the performance of page-table reads is negligible. Similarly, we see that Linux, Mitosis, and \sysname experience a similar rate of TLB shootdowns, indicating a similar overhead of page-table updates. As such, Mitosis and \sysname (without the TLB optimization) achieve  similar performance to Linux. However, as we can see in Figure Figure~\ref{fig:apache}b, \sysname with the TLB optimization incurs about 45\% reduction in TLB shootdowns. This reduction in TLB shootdowns results in about 18-20\% increase in throughput, as shown in Figure~\ref{fig:apache}a.

\subsubsection{Case Study: \textbf{In-memory key-value store.}}

In-memory key-value stores, such as Memcached \cite{memcached}, are widely used in storing keys and values in memory to achieve low latency and high throughput. To protect the data store from data leakage, sensitive information corruption, or arbitrary accesses, \texttt{mprotect} is applied to the critical data section to protect the stored data \cite{park2019libmpk, zhuo2021rearchitecting, gu2022epk}. We create 8 client threads that use libMemcached to send SET/GET requests. The proportion of SET and GET are 0.1 and 0.9. The size of the keys and values are 32B and 256B, respectively. The Memcached instances allow up to 1024 concurrent connections. The memory used for Memcached storage is 10GB in this experiment. To maximize the scalability, we employ a varying number of Memcached processes, with two threads per process. The threads are evenly distributed across four sockets.

Figure \ref{fig:memcached} shows the normalized Memcached throughput and TLB shootdown rate of Mitosis and \sysname over unmodified Linux. \sysname with TLB optimization always gains consistent speed-ups with a geomean improvement of 36\% across all thread counts, while Mitosis suffers a slowdown, \revised{due to the overhead of synchronously keeping all page replicas coherent upon every page-table modification}. The performance improvements of \sysname come from the reduction in TLB shootdowns, shown in Figure \ref{fig:memcached}. \sysname with TLB optimization reduces the occurrence of shootdowns by \revised{50\% to 96\%}.
\section{Discussion and Future Work}
\label{sec:future}

\noindent \textbf{Support for Transparent Huge Pages.} Transparent Huge Pages (THPs) are known to improve system performance by reducing the number of page-table levels that need to be traversed when performing a translation and increasing the amount of memory that can be mapped by a single TLB entry. However, THPs will still suffer from limited TLB reach in systems with large amounts of main memory. As a result, THPs would still benefit from the faster address translations that \sysname provides. Extending \sysname to support THPs is a relatively straightforward task as we would simply need to implement the same duplication and coherence mechanisms that we had previously implemented on the THPs. However, support for THPs was not considered a priority for this version of \sysname as prior work has shown that THPs may not be the best choice~\cite{lwn:THPproblems, redis:disableTHP, mongodb:disableTHP, percona:disableTHP, panwar:huge}, especially for NUMA systems~\cite{gaud:harmfulTHP, zi:nimble}. 

\noindent \textbf{Support for Virtualization.} Virtualized systems are another potential area for future work to explore. These systems typically make use of hardware-based nested paging \cite{gandhi:agilepaging, vmitosis2021} to translate the guest virtual address to the guest physical address and from the guest physical address down to the host physical address. Recent work has shown that virtualized systems are particularly sensitive to poor page-table placement~\cite{vmitosis2021}. While Mitosis has been extended to support 2D page-tables under vMitosis, vMitosis still works based on explicit policies given by the developers or providers. Unfortunately, it is challenging to maintain particular policies per applications and workloads. Furthermore, vMitosis also suffers from the same performance and memory overheads that Mitosis has since they both make use of eager replication. Thus, we believe that we will continue to see positive results once \sysname is extended to support the lazy replication of both the guest and host page-tables. 

\section{Related Work}
\label{sec:related}

In this section, other relevant works that have not been covered in earlier sections. The problem of remote page-tables in NUMA systems is an issue that has not received much attention. Despite an extensive search we were only able to find few prior works that directly highlighted and addressed the issue. However, there are other related works indirectly address the problem by either (1) minimizing the need for address translations or by (2) speeding up the address translation process.

\subsection{Segmented Address Spaces}
Corey~\cite{boyd2008corey} is an experimental operating system designed for many-processor systems. In Corey, the address space of each process is partitioned into two a shared region and a private one. While this method ensures that page-tables for the private partition are located locally, accesses to the shared partition might still result in a remote page-table access since the shared page-tables are not replicated. In addition, Corey requires the application to explicitly partition the address space which places an unnecessary burden on the programmer that makes unlikely for this technique to see widespread adoption.

\subsection{Minimizing the Need for Address Translations}
The related works below aim to reduce the need for address translations through various means, which can help reduce the performance penalties of remote pages by reducing the need to do page-table walks. However, as we see later on, these works are generally either (1) disruptive changes that compromise key functionalities of existing implementations of virtual memory or (2) incremental changes whose effectiveness is fundamentally limited by existing structures and systems.  

\textbf{Memory Segmentation} - Memory segmentation is a memory management technique that was used in early x86 machines and allowed the OS to allocate variable sized segments to programs. While it has long since fallen out of popularity in favour of paging, some recent works have decided to explore the idea as a possible way to extend the reach of TLBs \cite{basu2013efficient, karakostas2015redundant}. However, such methods prevent the use of mechanisms such as Copy-on-Write, demand-paging and per-page protection. Furthermore, these works are not robust to fragmentation as they require the segments to be allocated in contiguous blocks.

\textbf{Direct and Set-associative mappings} - Set-associative mappings restrict the range of physical addresses a particular virtual address can map to \cite{picorel2017near}. In more extreme cases, some of these methods directly map physical to virtual addresses \cite{haria2018devirtualizing}. While these methods virtually eliminate the overhead associated with address translation, they do not support important mechanisms like Copy-on-Write and demand paging in some cases \cite{haria2018devirtualizing}. In addition, such methods may also compromise the security of a system as it might impact the functionality of memory-protection mechanisms like address space layout randomization.

\textbf{Multi-page mapping} - Multi-page mappings allow translations with contiguous physical addresses to be mapped to a single entry in the TLB \cite{pham2012colt,   pham2014increasing}. While these methods do increase the reach of TLBs, the improvement is still not sufficient to support modern systems that have several hundreds of gigabytes of memory~\cite{gandhi2016range}.

\subsection{Speeding Up Address Translation}
The related works below seek to improve the speed of address translation in various ways and are largely orthogonal to our work.

\textbf{RadixVM}~\cite{clements2013radixvm} proposes a number of radical changes of the memory management, including lazy replication of memory management structures across all cores, which allows it to eliminate unnecessary TLB shootdowns, similarly to \sysname. Unfortunately, RadixVM is not scalable due to the per-core page-table replication, which results in huge memory overheads. For example, on our machine with 288 cores, around 60\% of the entire memory would be occupied by page-tables. RadixVM also requires a complete redesign of the kernel architecture and cannot be easily integrated into commercial OSes such as Linux. \textbf{LATR}~\cite{kumar:latr} modifies a number of memory management system calls to introduces lazy and batched issuance of TLB shootdowns and is orthogonal to our work.

\textbf{Prefetched Address Translation}~\cite{margaritov2019asap} takes advantage of the structure of page-tables to prefetch the lower levels of the page-table during a TLB miss. This reduces page-table walk latency without significant changes to existing systems. \textbf{Barrelfish}~\cite{barrelfish} relies on message passing instead of IPIs to shoot down TLBs in remote cores, eliminating inter-processor interrupts and lowering the latency of TLB shootdowns. Amit et al.~\cite{amit:dont} present a solution to accelerate the TLB shootdown performance by avoiding synchronous TLB shootdown calls mainly. Unfortunately, this approach is entirely oblivious to page-table replication design, and it won't be able to prevent extra TLB shootdown calls introduced by replicas. Nevertheless, \sysname can achieve even higher performance if it integrates such TLB shootdown acceleration solutions.

\section{Conclusions}
\label{sec:conclusion}
We have presented \emph{\sysname}, a novel on-demand partial page-table replication mechanism for NUMA systems that simultaneously optimizes both read and update accesses to page-tables. We have shown that page-table replication, if done lazily, can be leveraged to improve rather than degrade the performance of memory management operations. Using the extra knowledge of the sharers of a page-table afforded by our lazy replication, \sysname is also to reduce unnecessary TLB shootdowns and improve the runtime of memory management operations by up to 40x. At the same time, \sysname reaps all the benefits of page-table replication, providing a 20\% improvement over a wide range of applications, and provides scalability to any number of NUMA nodes.

% \section*{Acknowledgements}
% This document is an updated version of HPCA 2022 and 2023, which, in
% turn, has been derived from two previous conferences, in particular
% HPCA 2021 and MICRO 2021, which, in turn, are derived from past MICRO,
% HPCA, ISCA, and ASPLOS conferences.

%%%%%%%%%%%%%%%%%%%%%%%%%%%%%%%%%%%%
{ \bibliographystyle{plain}
\bibliography{refs}}
%%%%%%%%%%%%%%%%%%%%%%%%%%%%%%%%%%%%

\end{document}